# Superconductivity in iron-based F-doped layered quaternary compound Nd[O$_{1-x}$F$_x$]FeAs

Zhi-An Ren*, Jie Yang, Wei Lu, Wei Yi, Xiao-Li Shen, Zheng-Cai Li, Guang-Can Che, Xiao-Li Dong, Li-Ling Sun, Fang Zhou, Zhong-Xian Zhao*

National Laboratory for Superconductivity, Institute of Physics and Beijing National Laboratory for Condensed Matter Physics, Chinese Academy of Sciences, P. O. Box 603, Beijing 100190, P. R. China

**Abstract:**

Here we report a new quaternary iron-arsenide superconductor Nd[O$_{1-x}$F$_x$]FeAs, with the onset resistivity transition at 51.9 K and Meissner transition at 51 K. This compound has the same crystal structure as LaOFeAs, and becomes the second superconductor after Pr[O$_{1-x}$F$_x$]FeAs that superconducts above 50 K.

The recently discovered quaternary arsenide oxide superconductor La[$O_{1-x}F_x$]FeAs with the superconducting critical transition temperature ($T_c$) of 26 K [1], has been quickly expanded to another family of high-$T_c$ superconducting systems besides copper oxides by the replacement of La with other rare earth elements, such as Sm, Ce, and Pr *etc.* [2-4], where the Pr[$O_{1-x}F_x$]FeAs that synthesized under high pressure has become the first non-cuprate superconductor having a $T_c$ above 50 K. All these arsenide (including phosphide) superconductors formed in the same tetragonal layered structure with the space group P4/nmm that has an alternant stacked Fe-As layer and RO (R = rare earth metals) layer. Here we report the discovery of a new superconductor in this family, the neodymium-arsenide Nd[$O_{1-x}F_x$]FeAs with a resistivity onset $T_c$ of 51.9 K, which is the second non-cuprate compound that superconducts above 50 K.

The superconducting Nd[$O_{1-x}F_x$]FeAs samples were prepared by a high pressure synthesis method directly. Nd pieces, As, Fe, $Fe_2O_3$, $FeF_3$ powders (the purities of all starting chemicals are better than 99.99%) were mixed together according to the nominal stoichiometric ratio of Nd[$O_{0.89}F_{0.11}$]FeAs, then ground thoroughly and pressed into small pellets. The pellets were sealed in boron nitride crucibles and sintered in a high pressure synthesis apparatus under the pressure of 6 GPa and temperature of 1300°C for 2 hours. Here the adoption of high pressure synthesis is mainly due to the severe fluorine loss that we observed when using the common vacuum quartz tube seal method [1]. The press-seal of high pressure synthesis is better for synthesizing gas-releasing compound, while the disadvantage is that the short synthesis time and unstable condition are difficult for producing single-phase sample. The structure of the samples was characterized by powder X-ray diffraction (XRD) analysis on an MXP18A-HF type diffractometer with Cu-$K_\alpha$ radiation from 20° to 80° with a step of 0.01°.

The XRD patterns indicate that all samples include mixed phases, while the main phase adopts the same PrOFeAs structure as shown in Fig. 1. The impurity phases have been determined to be the known oxides, arsenides, and fluorides that are formed by starting chemicals, which do not superconduct at the measuring temperature. The existence of impurity phases is due to the insufficient synthesis time that can be tolerated by our high-pressure apparatus. Comparing with the Pr[$O_{0.89}F_{0.11}$]FeAs superconductor, the right-shifted XRD peaks indicate a shrinkage of lattice parameters for this Nd[$O_{0.89}F_{0.11}$]FeAs superconductor. A detailed study of the superconducting phase diagram will be reported later.

The resistivity was measured by the standard four-probe method. The results are shown in Fig. 2. A clear resistivity drop was observed as the temperature is down to 51.9 K, and the resistivity was unmeasurable below 48.8 K. The middle of the superconducting transition is at 50.1 K. Comparing with Pr[$O_{0.89}F_{0.11}$]FeAs superconductor, the $T_c$ (zero) increases about 5 K while the $T_c$ (onset) has no obvious change. Currently we believe that the chemical pressure caused by the shrinkage of crystal lattice promotes the increase of $T_c$ [5], which is indeed observed by our series of superconducting samples with different rare earth metal substitutions. This observation is also consistent with a theoretical expectation proposed in Ref. [6], where it is indicated that the $T_c$ may be enhanced by the increase of hopping integral, which can be achieved by the shrinkage of the lattice.

The magnetization measurements were performed on a Quantum Design MPMS XL-1 system during warming cycle under fixed magnetic field after zero field cooling (ZFC) and field cooling (FC) process. The AC-susceptibility data (with the measuring frequency of 997.3 Hz and the amplitude of 0.3 Oe) and DC-susceptibility data (measured under a magnetic field of 1 Oe) are shown in Fig. 3 (we note that the background is caused by magnetic impurities). The sharp magnetic transitions on both of AC and DC susceptibility curves indicate the good quality of this superconducting component. The onset diamagnetic transition determined from the differential ZFC curve is 51 K, which is a little higher than that of Pr[$O_{0.89}F_{0.11}$]FeAs superconductor. Since currently the best synthesis condition for this superconducting system is in exploring, a higher $T_c$ in this family is still expected.

Acknowledgements:

We thank Mrs. Shun-Lian Jia for her kind helps in resistivity measurements. This work is supported by Natural Science Foundation of China (NSFC, No. 10734120) and 973 program of China (No. 2007CB925002). We also acknowledge the support from EC under the project COMEPHS TTC.

Corresponding Author:

Zhi-An Ren: renzhian@aphy.iphy.ac.cn

Zhong-Xian Zhao: zhxzhao@aphy.iphy.ac.cn

Figure captions:

Figure 1: X-ray powder diffraction pattern of the Nd[$O_{0.89}F_{0.11}$]FeAs superconductor compared with that of Pr[$O_{0.89}F_{0.11}$]FeAs superconductor; the arrows indicate the main phase diffraction peaks and the vertical bars correspond to the calculated diffraction intensities.

Figure 2: The temperature dependence of resistivity for the Nd[$O_{0.89}F_{0.11}$]FeAs superconductor.

Figure 3: The temperature dependence of AC-susceptibility, DC-susceptibility, and differential ZFC curve for the Nd[$O_{0.89}F_{0.11}$]FeAs superconductor.

Figure 1:

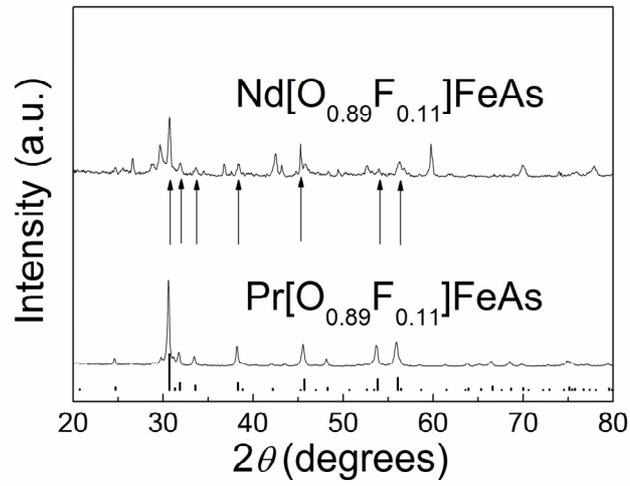

Figure 2:

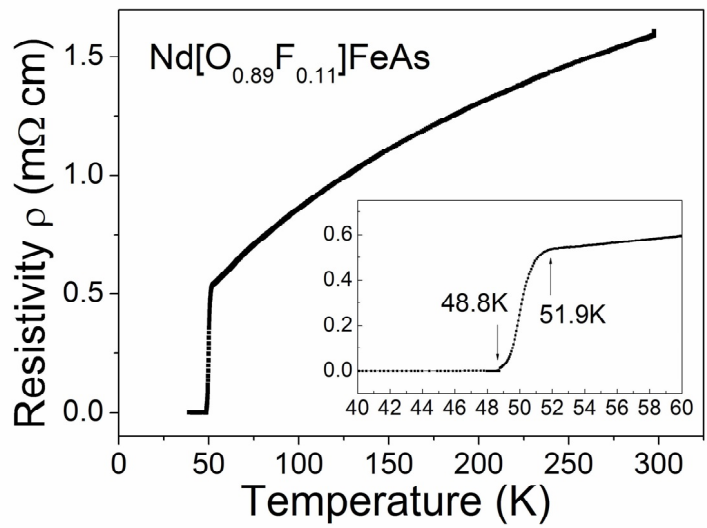

Figure 3:

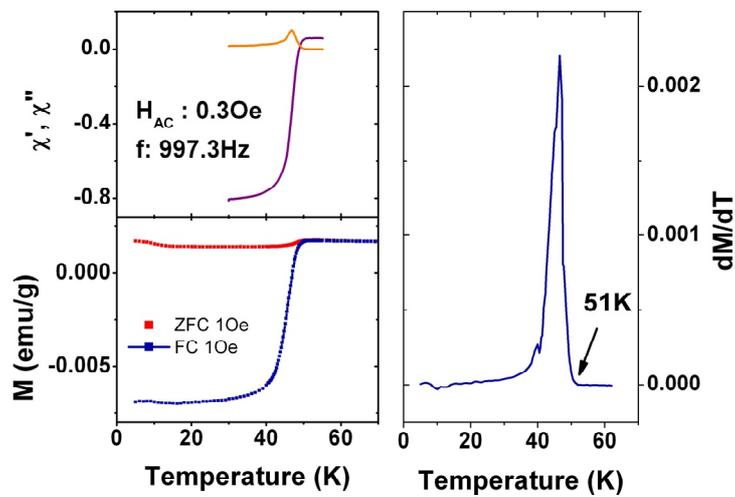

5